\documentclass[useAMS,usenatbib]{mn2e}

\usepackage{times}
\usepackage{amssymb}
\usepackage{graphicx}

\title[A transient QPO from GRS 1915+105]
{A transient low-frequency QPO from the black hole binary GRS 1915+105}
\author[P. Soleri, T. Belloni and P. Casella]{P. Soleri$^{1,2,3}$\thanks{E-mail:
psoleri@science.uva.nl}, T. Belloni$^{2}$, P. Casella$^{1,2}$\\
$^{1}$Astronomical Institute `Anton Pannekoek', University of
Amsterdam and Center for High Energy 
Astrophysics, Kruislaan 403, 1098 SJ, Amsterdam,\\ the Netherlands\\
$^{2}$INAF -- Osservatorio
Astronomico di Brera, Via E. Bianchi 46, I-23807 Merate, Italy\\
$^{3}$Universit\`a degli Studi di Milano, via Celoria 16, 20133 Milano,
Italy}

\begin{document}

\date{Accepted 2007 October 16. Received 2007 September 24; in original form 2007 July 20}

\pagerange{\pageref{firstpage}--\pageref{lastpage}} \pubyear{2007}

\maketitle

\label{firstpage}

\begin{abstract}
We present the results of the timing analysis of five
{\it Rossi X-ray Timing Explorer}
observations of the Black Hole Candidate GRS 1915+105
between 1996 September and 1997 December.
The aim was to investigate the possible presence of a type-B
quasi-periodic oscillation (QPO).
Since in other systems this QPO is found to appear during spectral
transitions from {\it Hard} to {\it Soft} states, we analyzed
observations characterized by a fast and strong variability, in order to have a
large number of transitions. In GRS 1915+105, transitions occur on very short
time scales ($\sim$ sec): to single them out we averaged Power Density
Spectra following the regular path covered by the source on a 3D
Hardness-Hardness-Intensity Diagram. 
We identified both the type-C and the type-B quasi-periodic
oscillations (QPOs): this is the first detection of a type-B QPO in
GRS 1915+105.  As the spectral
transitions have been associated to the emission and collimation of
relativistic radio-jets, their presence in the prototypical galactic
jet source strengthens this connection.
\end{abstract}

\section{Introduction} \label{par:intro}

Systematic variations in the energy spectra and intensity of transient
Black-Hole Candidates (BHC) have been recently identified in terms of
the pattern described in an X-ray Hardness-Intensity diagram (HID, see
Homan et al. 2001, Homan et al. 2005b, Belloni et
al. 2005). Four main bright states (in addition to the quiescent
state) have been found to correspond to different branches/areas of a
square-like HID pattern. In this framework much importance is given to
the intermediate states (called Hard Intermediate State, HIMS, and
Soft Intermediate State, SIMS) and to the transitions between them,
identified from the behaviour in several bands of the electromagnetic
spectrum (from radio to hard X-rays, see also Fender, Belloni \& Gallo
2004 and Homan et al. 2005b) and from the timing properties of the X-ray light
curve.

Low-frequency Quasi-Periodic Oscillations (LFQPOs) with centroid
frequency ranging from mHz to tens of Hz have been observed in the
X-ray flux of many galactic BHCs since the '80s (see van der Klis
2006; McClintock \& Remillard 2006 and references
therein). Three main types of LFQPOs, dubbed Type-A, -B and -C
respectively, originally identified in the light curve of XTE
J1550-564 (Wijnands et al. 1999; Remillard et al. 2002), have been
seen in several sources (see Casella et al. 2005 and references
therein). We summarize their properties in Table \ref{ABC_properties}.

In the context of the state classification outlined above, it is
possible to ascribe the three LFQPOs to different spectral conditions
(see Table \ref{ABC_properties}, Homan et al. 2001, Homan \& Belloni 2005,
 Belloni et al. 2005).

The type-C QPO is associated to the (radio loud) HIMS and to the
low/hard state. It is a common QPO seen in almost all BHCs with a
variable centroid frequency correlated with the count rate, a high
fractional variability and a high coherence ($Q=\nu/$FWHM$\sim$10).

The type-B QPO has been seen only in few systems, although it is being seen
in a growing number of sources (see Casella et
al. 2005 and references therein). It is a transient QPO associated to
spectral transitions from the (radio loud) HIMS to the (radio quiet)
SIMS. Its features are a $\sim$ fixed centroid frequency (around
$\sim$6 Hz), lower fractional variability and $Q$ than type-C. Some
authors (Fender, Belloni \& Gallo 2004; Casella et al. 2004) suggested that these
spectral transitions are in turn associated to the emission and
collimation of transient superluminal relativistic jets visible in radio
band. These jets are seen in a number of sources (GRS 1915+105, XTE
J1550-564, GX 339-4, XTE J1859+226, GRO J1655-40, etc.). However, not
in all of these sources we could resolve the spectral transition to
see the transient QPO.

\begin{table} \label{tab:ABC}
\centering
\caption{Summary of type-A, -B and -C LFQPOs properties (from Casella et
  al. 2005).}
\label{ABC_properties}
\scriptsize
\begin{tabular}{lccc}
\hline
\hline
Properties & Type-C & Type-B & Type-A \\
\hline
Frequency (Hz) & $\sim0.1-15$ & $\sim5-6$  & $\sim8$ \\
Q($\nu /$FWHM) & $\sim7-12$ & $\gtrsim6$ & $\lesssim3$ \\
Amplitude (\% {\it rms}) & 3-16 & $\sim2-4$ & $\lesssim3$ \\
Noise & strong flat-top & weak red & weak red \\
Phase lag$^{\mathrm{a}}$ @$\nu_{QPO}$ & soft$/$hard$^{\mathrm{b}}$ & hard & soft \\
Phase lag @2$\nu_{QPO}$ & hard & soft & ... \\
Phase lag @$\nu_{QPO}/2$ & soft & soft & ... \\
\hline
\end{tabular}
\begin{list}{}{}
\item[$^{\mathrm{a}}$] With ``hard lag'' we mean that hard variability lags
  the soft one.
\item[$^{\mathrm{b}}$] Trend towards soft lags for increasing QPO frequencies
\end{list}
\end{table}

The spectral properties connected to type-A QPO are similar to those
introduced for the type-B. This QPO has been seen in few systems
(Casella et al. 2005). It is broader, weaker and less coherent than
the type-B QPO.

GRS 1915+105 is a transient BHC discovered on August 15 1992 with the
WATCH instrument on board GRANAT (Castro-Tirado et al. 1992, 1994). It
is the first galactic source observed to have apparently superluminal transient
relativistic radio jets (Mirabel \& Rodriguez 1994), commonly
interpreted as ejection of ultra-relativistic plasma, with a speed
close to the speed of light (up to $\sim98$\%). Its radio variability
was discovered to correlate with the hard X-ray flux (Mirabel et
al. 1994). Thanks to VLA-radio observations of these jets, Rodriguez
et al. (1995) estimated a distance of 12.5 kpc; more
recent estimates attest a distance of $6.5 \pm 1.6$ kpc (Kaiser et al.
2005). The mass of the compact object was estimated through IR
spectroscopic studies (Greiner, Cuby \& McCaughrean 2001, Harlaftis \&
Greiner 2004) to be $14.0 \pm 4 M_{\odot}$ which unambiguously makes
GRS 1915+105 a BHC.

The various and rich phenomenology of this source was classified by
Belloni et al. (2000): they analyzed 163 RXTE observations, showing
that the complex behaviour of GRS 1915+105 can be described in terms
of spectral transitions between three basic states, A, B and C (not to
be confused with the name of the LFQPOs introduced before), that give
rise to 12 variability classes. The non standard behaviour of GRS
1915+105 (it is a very bright transient source continuing the same
outburst started in 1992) was interpreted as that of a source that
spends all its time in Intermediate States (both in its hard and soft
flavors), never reaching the LS or the quiescence (see e.g. Fender \&
Belloni, 2004).

GRS 1915+105 also shows strong time variability on time scales of
fractions of second, revealing low- and high-frequency QPOs (see
Morgan et al. 1997) whose properties (frequency and fractional
variability) are tightly correlated with the spectral parameters
(Morgan et al. 1997; Muno et al. 1999; Markwardt et al. 1999;
Rodriguez et al. 2002a, 2002b; Vignarca et al. 2003). In particular,
all LFQPOs observed from this system can be classified as type-C
QPOs. Although GRS 1915+105 makes a large number of fast state
transitions, which have been positively associated to radio activity
and jet ejections, no type-B QPO has been observed to date.

\begin{table*}
\centering
\caption{Log of the 5 RXTE/PCA observations analyzed in this work} \label{log_obs}
\begin{tabular}{c c c c c c}
\hline
\hline
N$^{\circ}$ & Obs. Id.& Date & Starting MJD & Exp. (s) & Classification\\
\hline
1 & 10408-01-35-00 & 1996 Sep 09 & 50348.271 & 9448& $\mu$\\
2 & 20402-01-45-03 (\#1, 2, 3) & 1997 Sep 09 & 57700.250 & 10038& $\beta$\\
3 & 20402-01-53-00 & 1997 Oct 31 & 50752.013 & 9656& $\beta$\\
4 & 20402-01-53-01 \& 20402-01-53-02(\#1) & 1997 Nov 04-05 & 50756.412 & 6322&$\mu$\\
5 & 20402-01-59-00 & 1997 Dec 17 & 50799.091 & 9784& $\beta$\\
\hline
\end{tabular}
\flushleft
Obs. 4 is composed of two orbits from two separate observations\\
\# indicates the number of the RXTE orbit within the observation\\
Classification is from Belloni et al. (2000).
\end{table*}

In this paper we present the discovery with RXTE of the type-B QPO in
the X-ray light curve of GRS 1915+105. The QPO was present during fast
spectral transitions that we identify with the HIMS to SIMS transition
observed in other BHCs.

\section{Observations and data analysis} \label{par:obs_data}

\begin{figure}
\begin{tabular}{c}
\resizebox{7.7cm}{!}{\includegraphics{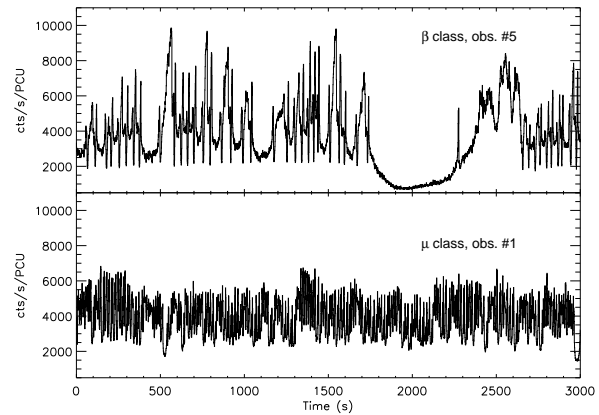}}
\end{tabular}
\caption{2-13 keV light curves for the two variability classes examined here. 
Bin size is 1 second.}
\label{fig:mu_beta_class}
\end{figure}

We analyzed five RXTE/PCA observations GRS 1915+105 collected during
AO1 and AO2, between 1996 September 22 and 1997 December 17 (see Table
\ref{log_obs}). We chose observations belonging to variability classes
$\mu$ and $\beta$ (according to the classification of Belloni et
al. 2000, see Figure \ref{fig:mu_beta_class} for two examples of light
curves). Restriction to AO1
and AO2 is due to three reasons: first, AO1 and AO2 observations are those for which a
thorough classification was made by Belloni et al. (2000), which  removes the need for
a complete classification of observations throughout the archive, a task beyond
the scope of the present paper; second, AO1 and AO2 were in average characterized
by a higher number of working PCUs compared to more recent epochs (this improves the
statistics) and third, using data from different epochs, you get different
hardness-intensity and color-color diagrams. We chose class $\mu$ and $\beta$ observations
because they correspond to intervals when frequent fast transitions among
the three spectral states are observed.

\begin{figure*}
\begin{tabular}{c}
\resizebox{15cm}{!}{\includegraphics{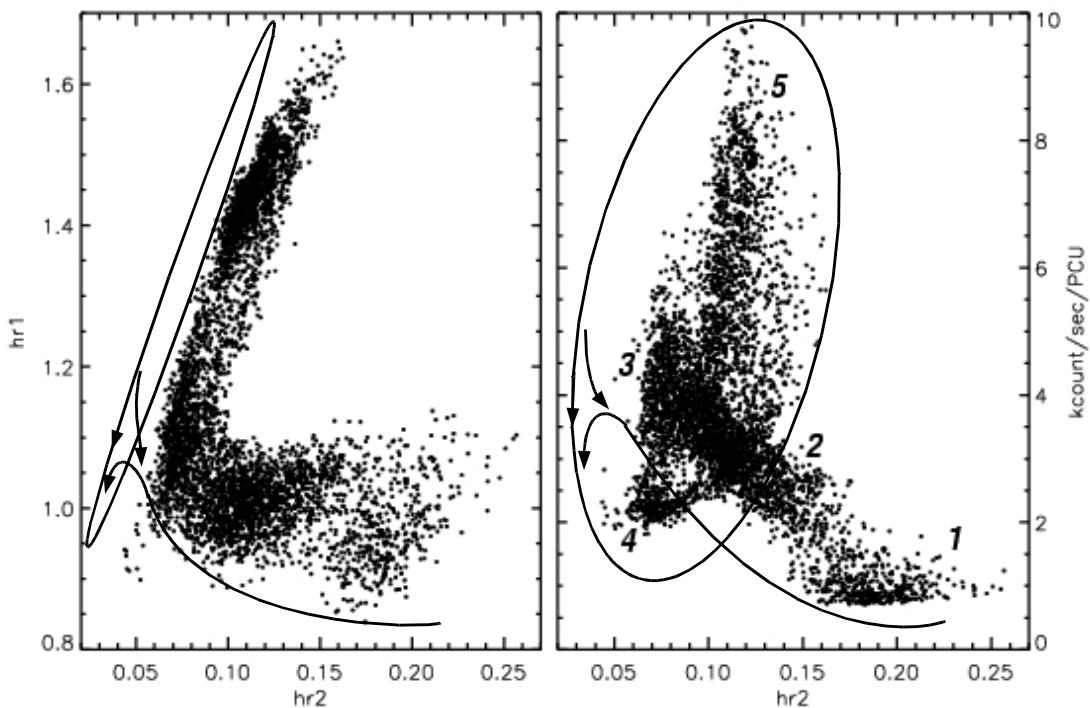}}
\end{tabular}
\caption{Two-dimensional projections of the hardness-hardness-intensity
diagram (HHID) for observation \#5. Left side panel: color-color
diagram; right side panel: hardness-intensity diagram. Definitions for
the colors can be found in \S\ref{par:obs_data}, the count rate is the
2-13 keV energy band. The arrows follow the source's regular and
repetitive movement. Numbers correspond to the regions discussed in the text (see
Figure \ref{fig:zone_QPO}).}
\label{fig:cd_hid_arrows}
\end{figure*}

The rationale behind this choice was to analyze a large number of
spectral transitions from hard (C) to soft states (A and B), when a
type-B QPO would be expected. The transitions being fast, the QPO
would not be detected for a single transition, but adding a large
number of data corresponding to the same transition, the signal could
become detectable. The total amount of usable exposure was 44648
seconds. For each observation we produced light curves in the PCA
energy channels 0-35 (corresponding to the energy range 2-13 keV) and
two color curves defined as the ratio between the counts in the 5-13
and 2-5 keV bands (HR1, absolute channels 14-35 and 0-13) and between
the counts in the 13-38 and 2-5 keV bands (HR2, absolute channels
36-103 and 0-13). The bin size was 2 seconds for all these curves.
Given the high count rates we did not subtract the background. In
addition, for each 2-second time interval we accumulated a power
density spectrum (PDS) in the 2-5 keV, 5-13 keV and 13-38 keV energy
ranges. All the PDS were extracted with a time resolution of $1/128$ s
(corresponding to a Nyquist frequency of 64 Hz).
The PCA data modes
used to produce the PDS were Binned and Event Data. For an explanation
of the different PCA modes see for example Jahoda et al. (1996).

For each observation, we created a 3-dimensional
hardness-hardness-intensity diagram (HHID), where the source follows a
regular path. For each observation we averaged PDS on a variable
number of points ($\sim$100-800, to have enough statistics), following
the whole diagram. The selections were made using the free software
for multi-variable management XGOBI ({\em
http://www.research.att.com/areas/stat/xgobi/}). For each selection
we rebinned the average PDS logarithmically, and the Poissonian noise,
including the Very Large Events (VLE) contribution (Zhang 1995, Zhang
et al 1995), was subtracted. The PDS were normalized to square
fractional {\it rms} (see Belloni \& Hasinger 1990) and fitted with a
combination of Lorentzians (see Nowak 2000; Belloni et al. 2002). In a
few cases (all the PDS of observation \#1 and one PDS of observation
\#3) the Poissonian subtraction left flat residuals at high
frequencies, hence we decided to account for it adding an additive
constant to the model. The fitting was carried out with the standard
XSPEC v11.3 fitting package, by using a one-to-one energy-frequency
conversion and a unit response.

For each 2-s interval, we also produced a cross-spectrum between the
2-4.5 keV (absolute channels 0-11) and 4.5-11 keV (absolute channels
12-29) light curves. We then calculated averaged cross-spectrum
vectors for all the different selected regions of each observation,
from which we derived phase lag spectra (for details on the phase lag
analysis see e.g. Casella et al. 2004). We defined phase lags as
positive when the hard X-ray variability {\it follows} the soft
one. In order to quantify the phase-lag information for each detected
QPO component we integrated the lags in a range centered on the QPO
centroid frequency with a width equal to the {\em FWHM} of the QPO
peak itself (see Reig et al. 2000). Notice that the estimate of the
QPO lags depends on the relative power of the continuum and of the QPO
itself. When the continuum power contribution is significantly larger
than the QPO power the lags of the continuum may dominate.

\section{Results} \label{par:results}

\begin{figure*}
\begin{tabular}{c}
\resizebox{16cm}{!}{\includegraphics{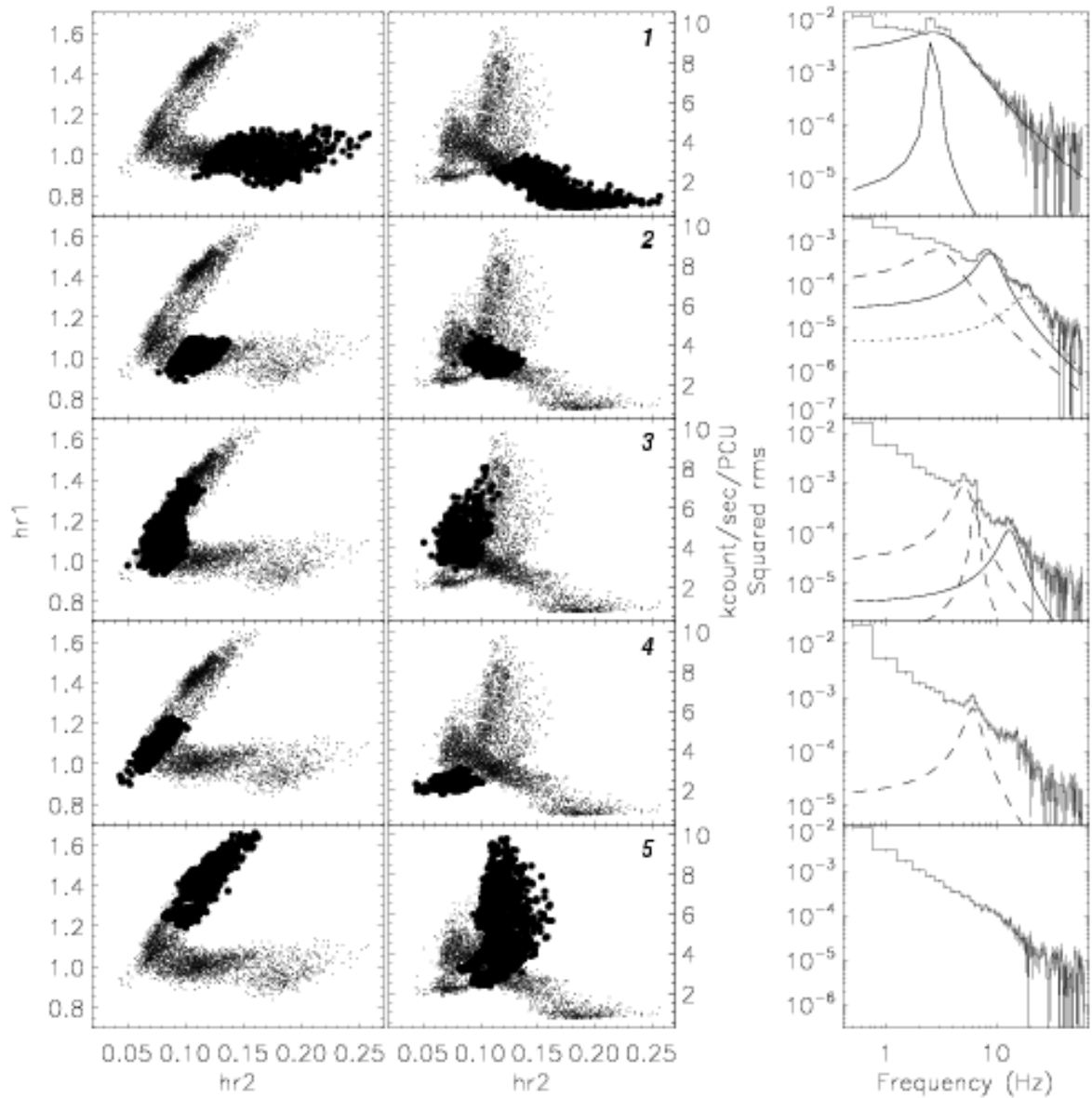}}
\end{tabular}
\caption{{\it Left panels:} Two-dimensional projections of the HHID
(see Fig. \ref{fig:cd_hid_arrows}). The tick points mark the five main
regions.  {\it Right panel:} Examples of five corresponding typical
power density spectra. For each power spectrum we also plot the
best-fit Lorentzian components corresponding to the significant
QPOs/bumps ($\sigma \geq 3$). Dashed lines: type-B QPOs, continuous
lines: type-C QPOs, dotted lines: harmonics. In the PDS in region {\em
4} the Lorentzian fitting a bump identified as the harmonic of the
type-B QPO ($\nu=13.84$ Hz) has not been reported since it has been
detected with a significance $\sigma<3$.}
\label{fig:zone_QPO}
\end{figure*}  
\begin{figure}
\begin{tabular}{c}
\resizebox{8cm}{!}{\includegraphics{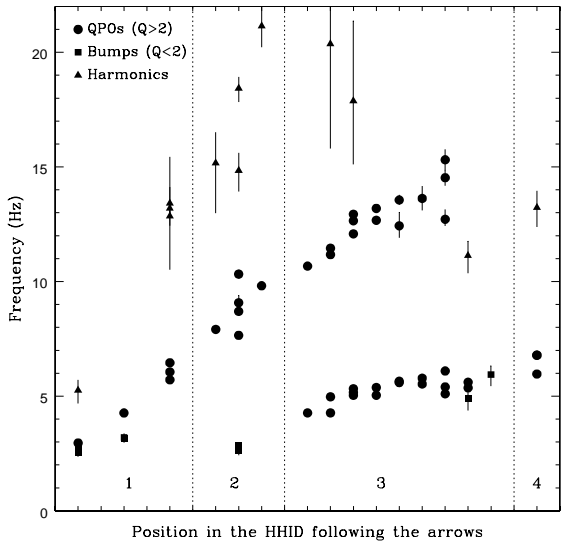}}
\end{tabular}
\caption{Evolution of QPOs/bumps (with significance $\sigma \geq 3$) centroid
frequencies as a function of the position in the HHID in Figure
\ref{fig:zone_QPO}, for all the analyzed observations. Adopted criteria are
explained in the text.}
\label{fig:freq_reg}
\end{figure}  
\begin{table}
\centering
\caption{QPO types detected in the five regions for all considered observations.
Region 1 is not present in observations \# 2 and 3. We indicated with
``Bump-B'' peaks with $Q<2$.} \label{tab:zone}
\begin{tabular}{c c c c c c}
\hline
\hline
  & \multicolumn{5}{c}{Regions in figure \ref{fig:zone_QPO}}\\
Obs. \# & 1 & 2 & 3 & 4 & 5 \\
\hline
1 &  -  & Bump-B, C & C, B & B        & No peaks \\
2 &  C  & C	    & C, B & No peaks & No peaks \\ 
3 &  C  & Bump-B, C & C, B & No peaks & No peaks \\
4 &  -  & C	    & C, B & B        & No peaks \\
5 &  C  & Bump-B, C & C, B & B        & No peaks \\ 
\hline
\end{tabular}
\end{table}

An example of HHID is shown in the two-dimensional projections in
Fig. \ref{fig:cd_hid_arrows} (from observation \#5, class
$\beta$). The regular and repetitive movement followed by the source
is marked by the arrows. Starting from the bottom-right corner of both
projections, the PDS properties were found to change smoothly thorough
the diagram. We could identify five main average behaviours of the
PDS corresponding to five regions in the diagram. In Figure
\ref{fig:zone_QPO} we show the five regions in the two-dimensional
projections of the HHID of observation \#5, together with five
corresponding examples of PDS. In Figure \ref{fig:freq_reg} we show
the evolution of the frequencies of all significant peaks (with
significance $\sigma \geq 3$) for all the observations, moving through
the HHID following the arrows in Figure \ref{fig:cd_hid_arrows}. The x axis has been
chosen in such a way to have an equal spacing between different selections in the
same region, since the same region in different observations has been divided in a
different number of selections. This point can be clarified by looking tables
\ref{tab:obs1} to \ref{tab:obs5} in appendix \ref{app_tables}, where we
reported all the significant peaks different
from a continuum fitted in all the selections of the five observations.
For each selection (in figure \ref{fig:freq_reg}) we reported the peak frequency
obtained from the energy
range where the feature was more significant. In four cases we found
that the same feature was detected at more than 3 sigma as a QPO
(i.e. with a quality factor $Q\geq 2$) in one energy band and, with a
higher significance, as a bump ($Q<2$) in another. In these cases we
decided to report the frequency obtained from the QPO. We considered
two peaks as harmonics of each other if (i) their centroid
frequencies, fitted independently, are consistent with being
harmonically related or (ii) fixing their centroid frequency to an
harmonic ratio we could obtain a very good fit. We will no longer
consider harmonic peaks in our discussion.

In Figure \ref{fig:zone_QPO} we can easily identify the presence of
two peaks evolving in frequency thorough the diagram from region {\em
1} to region {\em 4} (see Table \ref{tab:zone} for a summary of their
properties):

\begin{enumerate}
   \item the first peak is detected (superimposed to a broad-band
   noise) since region {\em 1}. It increases in frequency (from
   $\sim$2 Hz to $\sim$15 Hz) and quality factor $Q$ (from 0.64
   to 7.39) and decreases in fractional {\it rms} amplitude (from
   22.86 to 1.16) while moving thorough the diagram from
   region {\em 1} to {\em 2} until it disappears at high count rates
   at the end of region {\em 3}. Its fractional {\it rms} is higher in
   the hardest band (13-38 keV);
   \item a second peak is detected in region {\em 2} (in three
   observations, see table \ref{tab:zone}) with properties markedly
   different from the first one. It has initially (in region {\em 2})
   a frequency between 2.44 Hz and 2.84 Hz and a quality factor $Q<2$.
   In regions {\em 3} and~{\em 4} it shows a much less variable
   frequency than the first peak (slowly increasing from $\sim$4 to 7
   Hz) and a $Q>2$. Its fractional {\it rms} is higher in the hardest
   band (13-38 keV), is lower than that of the first peak and is
   independent on the centroid frequency (see Fig. \ref{fig:rms_1}).
\end{enumerate}

All five analyzed observations follow the average behaviour
described above, with three main exceptions:

\begin{figure*}
\begin{tabular}{c}
\resizebox{17.5cm}{!}{\includegraphics{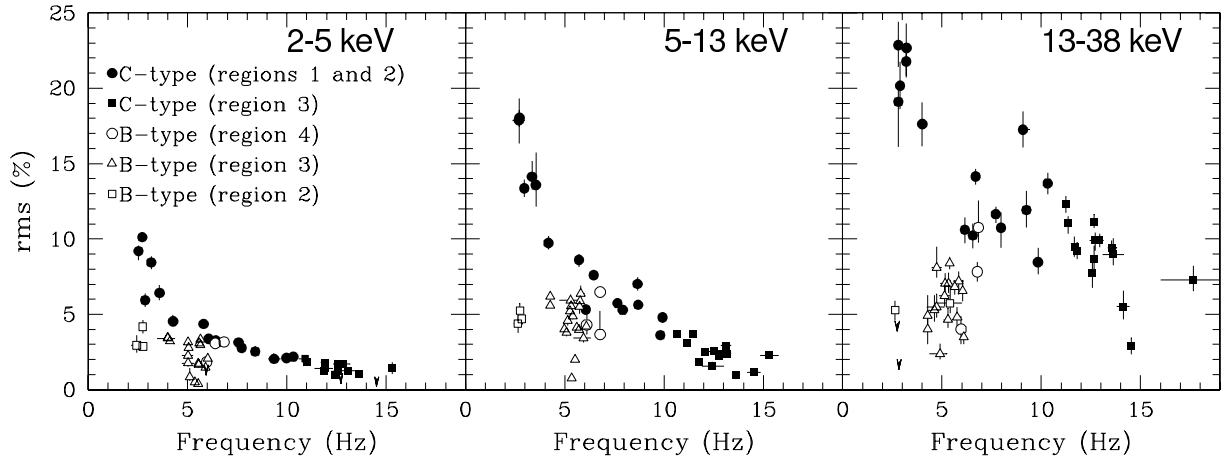}}
\end{tabular}
\caption{Fractional {\em rms} of the detected Bumps/QPOs in all observations
as function of their frequency for energy bands 2-5 keV (PCA channels
0-13, {\it left panel}), 5-13 keV (PCA channels 14-35, {\it center
panel}) and 13-38 keV (PCA channels 36-103, {\it right panel}).  Error
bars are estimated at 1 $\sigma$ confidence level. Arrows indicate 3
$\sigma$ upper limits for type-C QPOs only. Filled symbols indicate type-C QPOs,
open symbols type-B.}
\label{fig:rms_1}
\end{figure*} 

\noindent (a) region {\em 1}, which correspond to the long hard
intervals in the light curve of class $\beta$ (see
Fig. \ref{fig:mu_beta_class}, upper panel), are not present in the two
class $\mu$ observations (\#1 and \#4). Since this is the main
macroscopic difference between the two classes, and since a detailed
analysis of the long hard intervals of class $\beta$ light curves is
beyond the scope of this work, in the following of the paper we will
consider the two classes as a single class.

\noindent (b) No QPO is detected in the PDS from region {\em 4} of
observations \#2 (class $\beta$) and \#3 (class $\mu$). In region
{\em 4} of observation \#3, only in the highest energy band, we find a
non-significant QPO ($\sigma$=2.4) with the following properties:
$\nu$ = 5.85$\pm$0.26 Hz, $Q$ = 3.21$\pm$2.43, {\it rms} = (6.26$\pm$1.35)\%.
Both the {\it rms} amplitude value and the centroid frequency  are consistent
with those of the peak found in region {\em 4} in the 36-103 keV energy band
for observations \#1, 4 and 5, but the uncertainty on the quality factor $Q$
and the low significance ($<$3 $\sigma$) does not allow us to claim the real
detection of a QPO.

\noindent (c) In region {\em 2}, we detected two significant peaks in observations
\# 2, 3 and 5, while just one significant peak in observations \#2 and 4. In region
{\em 2} of observation \#4, we find a non-significant QPO  ($\sigma$=2.44) with
the following properties (values from 14-35 keV energy band):
$\nu$ = 3.46$\pm$0.18 Hz, $Q$ = 2.44$\pm$1.03, {\it rms} = (3.68$\pm$0.93)\%. Its
properties are consistent with those of the low-frequency peak found in region {\em
2} for observations \#1, 3 and 5, but the significance ($<$3 $\sigma$) does not
allow us to claim the real detection of a QPO. 

We will refer to the peak with centroid frequency ranging from $\sim$2
up to 15 Hz as {\it type-C} QPO and to the peak with 2.44--7 Hz frequency
detected in regions {\em 2}, {\em 3} and {\em 4} as {\it type-B} QPO. This
identification will be motivated below. In Figure \ref{fig:rms_1} we
show the fractional {\it rms} amplitude of all the detected Bumps/QPOs as a
function of their centroid frequency, for each of the three energy
bands analyzed. In a few cases QPOs were detected only in one or two
energy ranges. In these cases in order to obtain a more reliable
estimate of the peak amplitude also in the energy range where it was
not significant we fixed its centroid and FWHM to the values obtained
from the energy range in which it was more significant.

The points with large errors in frequency correspond to detections
where we need two Lorentzian components to obtain a satisfactory
fit. For these points we plot the average of the two centroid
frequencies with an error estimated as the sum between the two errors (we
considered this the most conservative approach possible).
In order to check whether these double peaks are due to the frequency
variability of the QPO we tried to select smaller regions in the
HHID. However we could not detect the QPO in such selections, due to
the low statistics of the resulting PDS.

\begin{figure*}
\begin{tabular}{c}
\resizebox{15cm}{!}{\includegraphics{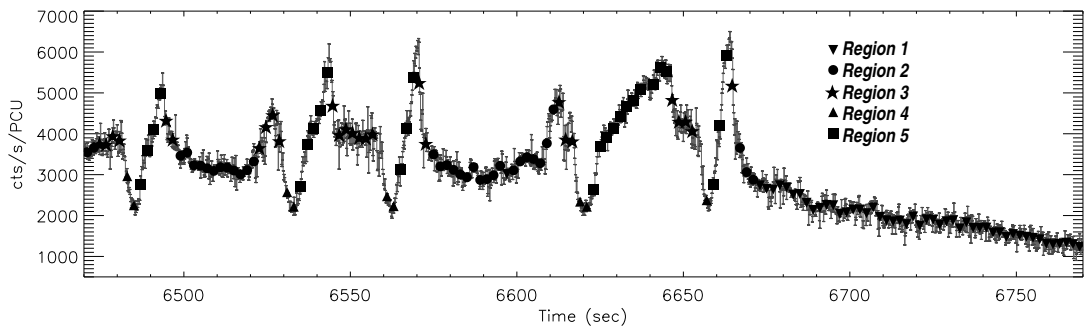}}
\end{tabular}
\caption{300-second light curve (2-13 keV energy band) for observation \# 5. Bin
size is 0.125 sec. Every symbol is averaged on 2 seconds. Different symbols
correspond to different regions in the HHID in Figure \ref{fig:zone_QPO}.}
\label{licu_symbols}
\end{figure*}

From Figure \ref{fig:zone_QPO} we can relate the presence of the
QPO-types to different zones of the source path in the HHID but we do
not have any detailed information about the time evolution of the
source position in that diagram. We know that $\mu$ and $\beta$
classes are characterized by fast and strong variability and frequent
spectral transitions but we do not know precisely between which
spectral states the transitions happen and when. A light curve where
parts corresponding to different regions in the HHID are marked with
different symbols can help us to clarify this point (see Figure
\ref{licu_symbols}). We chose a 300-second interval representative of
the source behaviour. With the identifications Downward Triangles =
Reg. {\em 1}, Circles = Reg. {\em 2}, Stars = Reg. {\em 3}, Upward
Triangles = Reg. {\em4} and Squares = Reg. {\em 5} we have the
following regions sequence: 2345 345 532. Class $\beta$ observations
are also characterized by the presence of long hard intervals (region
{\em 1}), occasionally occurring when the source maps the hard tail in
the bottom-right part of both the diagrams in Figure
\ref{fig:cd_hid_arrows}, always between region-{\em 2} intervals. If,
using the state classification presented in Belloni et al. (2000), we
make the identifications Reg. {\em 2} = C-state, Reg.  {\em 4} =
A-state, Reg. {\em 5} = B-state and we consider Reg. {\em 3} as a
Transition (T) state, we have the following state sequences: C T AB T AB
T C. In this framework region {\em 1} is the hardest part of the
C-state.
\subsection{Phase lags}
Phase lags for all the detected peaks are reported in Tables \ref{tab:lagall1}
and \ref{tab:lagall2} in appendix \ref{app_tables}. In Figure
\ref{fig:lag_C} we show phase lags vs. frequency for the
type-C QPO in two observations. If the QPO was fitted with two
Lorentzians we estimated the phase lag integrating on a peak centered
on the centroid frequency of the Lorentzian with the highest {\it rms}
amplitude and with the FWHM of the same component.  For observation \#
1, a negative trend for increasing QPO frequencies is visible,
although one point is completely out of this trend. For observation \#
5 the phase lag of the QPOs are always negative and consistent with
being constant.

\begin{figure}
\begin{tabular}{c}
\resizebox{8cm}{!}{\includegraphics{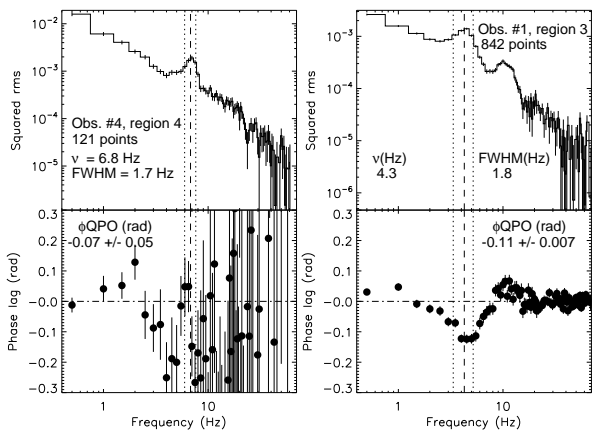}}
\end{tabular}
\caption{Power density spectra and phase-lag spectra of two different
observations/regions where a type-B QPO was detected.  The dashed
line marks the QPO frequency, the dotted lines mark $\nu \pm
HWHM$. Left panel: type-B QPO detected averaging power spectra (121
points) in region 4 of observation \# 5.  Right panel: type-B QPO
detected averaging power spectra (842 points) in region 3 of
observation \# 1.}
\label{fig:lag_B}
\end{figure}

In Figure \ref{fig:lag_B} we show two examples of power density
spectra (upper panels) and phase-lag spectra (bottom panels) from regions
where a type-B QPO was detected. In the left panel (observation \#4),
even thought the errors on phase lags are large, a turn towards
positive lag values in correspondence of the QPO frequency is
apparent. However, extracting the phase lags over the whole peak width
(i.e. over the frequency range $\nu_{p}\;\pm\;${\small FWHM/2}) we
obtain a negative value of the lags (although consistent with zero,
$\phi$ = -0.07$\pm$0.05 rad). In the right panel (observation \#1)
we have a much better statistics and the phase lags are clearly
negative ($\phi$ = -0.11$\pm$0.007 rad) in correspondence of the
type-B QPO (when integrated over the whole peak width).

\begin{figure}
\begin{tabular}{c}
\resizebox{7cm}{!}{\includegraphics{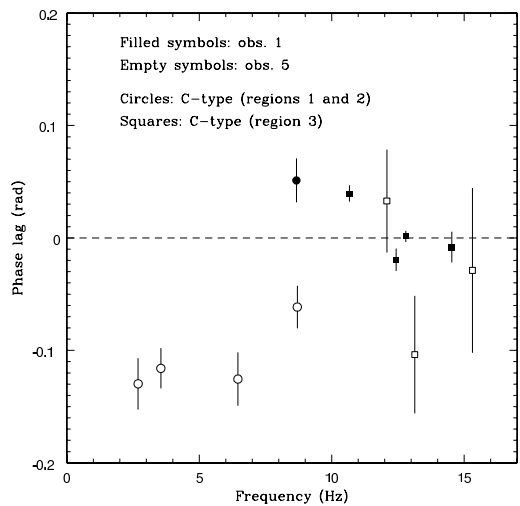}}
\end{tabular}
\vspace{-2.0cm}
\caption{Phase lags of type-C QPOs detected
in two representative observations. For each QPO we extracted the
phase lag in a range centered at the QPO peak frequency and
corresponding to the width itself ($\nu_p \, \pm \, FWHM/2$). Errors
bar on the X-axis are not shown for clarity.}
\label{fig:lag_C}
\end{figure}

\section{Discussion and conclusions} \label{par:discussion}


We analyzed 5 RXTE/PCA observations collected during the first two
years of the mission. In the power density spectra of all five
observations we detected several peaks which we identify with two
different types of QPO already seen in many other BHCs: the type-C and
type-B QPOs. This is the first identification of a type-B QPO in GRS
1915+105. To detect it, we looked in detail at spectral transitions in
observations characterized by a fast and intense variability. Spectral
transitions in GRS 1915+105 are usually very fast, often occurring on
timescales of $\sim$ seconds. This is at variance with most of other
black-hole binaries in which spectral transitions are observed to last
hours or days. In order to study the spectral transitions in GRS
1915+105 we performed an energy-dependent timing analysis by averaging
power spectra on the pattern the source recursively tracks in the
3-dimensional hardness-hardness-intensity diagram.

Applying this method, we found a type-B QPO in all five
observations. In all of them, we detected the type-B QPO together with
the type-C, in region \# 3 of the HHID (see Tab.  \ref{tab:zone}). In
three of them, we also detected a type-B QPO alone, in region \# 4 of
the HHID. Two of these observations belong to class $\mu$ and one to
class $\beta$, which excludes any relation between the class of
variability and the presence of the type-B QPO. This means that the
presence of the long hard intervals (which differentiate the
$\beta$-class from the $\mu$-class light curves) does not influence
the fast timing properties of the source outside these intervals.  We
could not find any property (as e.g. hardness, rate) correlated with
the presence of the type-B QPO alone in region {\em 4} of the HHID. In
three observations we also found a type-B bump in region {\em 2}. No correlations
were found neither between the presence of the
type-B bump and the type-B QPO in region {\em 4} nor with the hardness
and the count rate.

\begin{figure}
\begin{tabular}{c}
\resizebox{7cm}{!}{\includegraphics{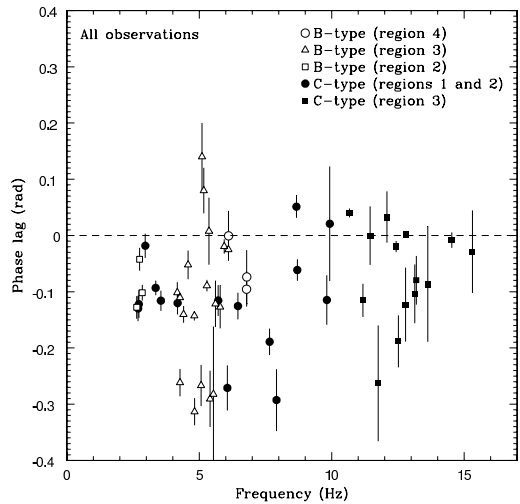}}
\end{tabular}
\vspace{-2.0cm}
\caption{Phase lags of the detected QPOs in all observations. For each
QPO we extracted the phase lag in a range centered at the QPO peak
frequency and corresponding to the width itself ($\nu_p \, \pm \,
FWHM/2$). Errors bar on the X-axis are not shown for clarity.}
\label{fig:lag_all}
\end{figure}

\subsection{QPOs identification} \label{subpar:disc_identif}

In order to identify the two types of QPO that we found in our data
sets, we compare them with the known LFQPOs in BHCs. In particular we
first analyze their position and behaviour in the Hardness-Intensity
diagram (right panel of Figure \ref{fig:cd_hid_arrows}). When the
source moves through regions \#{\em 1} and \#{\em 2} up to region
\#{\em 3} the first QPO shows a behaviour very similar to that of
type-C QPOs: its frequency is correlated with the count rate and is
inversely correlated with the hardness. At a certain hardness, this
QPO disappears. A second type of QPO is also detected: as in the case
of type-B QPOs, this second QPO appears in a narrow frequency range
(often around $\sim$6 Hz) and in a limited range in hardness. 

Its frequency and quality factor appear to be slightly correlated with
the count rate, particularly when at its lowest frequencies (2.44-2.84
Hz). We interpret this as the first evidence of an increase of the
coherence of the type-B QPO from $Q<2$ when at low frequencies to
$Q>2$ when reaching frequencies around $\sim$6 Hz, possibly suggesting
the presence of a resonance at this frequency (see Casella et
al. 2004).

The combined evolution of the QPOs in GRS 1915+105 is strongly
reminiscent of the known behaviour of type-C and type-B QPOs in BHCs
(see Casella et al. 2005 and references therein).

To verify this identification, we plot in Figure \ref{fig:rms_1} the
{\it rms} fractional variability of the detected QPOs as a function of
their frequency for three energy bands. The two QPO types have a
somewhat similar energy dependency (being stronger at high energy) but
they clearly show different behaviours in these diagrams. In each of
the three panels, two well-identified groups of points are evident. A
comparison of these two groups with Figure \#3 of Casella et al. 2004
helps to classify the observed QPO in GRS 1915+105: the first group is
diagonally spread across the plots, covering the whole frequency range
between $\sim2$ and $\sim15$ Hz and a large range in {\it rms} (particularly
at high energies, see the right panel of Figure \ref{fig:rms_1}). The
second group is clustered both in frequency (between $\sim2.5$ and
$\sim7$ Hz) and in fractional {\it rms}. The observed behaviour is clearly
consistent with that known to be typical of type-C and type-B QPOs in
BHCs (see Casella et al. 2004, 2005).

\subsection{Phase lags} \label{subpar:disc_lag}

The association between QPO-type and phase lag in literature is not
conclusive: although an average behaviour can be identified (see
Tab. \ref{tab:ABC}, Casella et al. 2005 and reference therein) there
are a number of exceptions (see e.g. Belloni et al. 2005 and Homan et
al. 2005a). Nevertheless we performed a phase-lag analysis in order to
have a comprehensive view of the behaviour of the QPOs in GRS 1915+105.

On the basis of the analyzed data it is not possible to characterize
unambiguously the phase lag behaviour of any of the two types of QPO
we observe: lags appear to vary between different observations,
although both types show in average negative values of phase lag (see
Fig. \ref{fig:lag_all}).

This can possibly due to the fast variability of the analyzed light
curves: we extracted power spectra over time intervals 2 seconds long,
without applying any procedure to detrend the variability on longer
time scales. However, this variability is very strong (see
Fig. \ref{fig:mu_beta_class}), which results in a leakage at higher
frequencies as strong as to actually dominate the phase-lag continuum.

\subsection{GRS 1915+105 as a ``normal'' source} \label{subpar:disc_normal}

The type-B QPO was detected in a few BHCs (see Casella et al. 2005),
and associated to spectral transitions from HIMS to SIMS (Homan \&
Belloni 2005, Belloni et al. 2005 and reference therein). Some authors
(Gallo et al 2003; Fender, Belloni \& Gallo 2004) suggested a relation
between these spectral transitions and the emission and collimation of transient
relativistic radio jets: if we consider the type-B QPO as the
signature of radio jet emission, we have this ``signature'' also in
the prototypical galactic jet source.  As already pointed out by
Belloni et al. (2005), the non-detection of a type-B QPO in GRS
1915+105 was rather interesting, especially if you interpret the
X-ray$/$radio correlation in this source and in other transients in
the framework of the same model (Fender, Belloni \& Gallo 2004). In
GRS 1915+105 the association between X-ray and radio activity is well
known (Pooley \& Fender 1997, Mirabel et al. 1998, Fender \& Belloni
2004 and references therein). Belloni et al. 2005 suggested that the
elusiveness of this QPO (in GRS 1915+105) could be due to high
velocity of the movement of the source through the HID. The result
presented in this work thus strengthens the interpretation of GRS
1915+105 in the framework of the same model of other BHCs: a type-B
QPO appears in correspondence of spectral transition from the B-state
to the A-state (Transition, Region {\em 3}, Stars in Figure
\ref{licu_symbols}) and when the source is in the A-state (Region {\em
4}, Upward Triangles in Figure \ref{licu_symbols}). In the light curve
in Figure \ref{licu_symbols} we see state oscillations CTAB TAB TAB
that we can identify as fast
passages between the HIMS and the SIMS of the other BHCs (Casella et
al. 2004; Fender, Belloni \& Gallo 2004). The type-B QPO appears also
in correspondence of the transition from the B-state to the C-state
(Squares-Stars-Circles in Figure \ref{licu_symbols}),
therefore not in an oscillation event but in a transition from a soft
to a hard state.

Furthermore, GRS 1915+105 is at present the heaviest black hole
(although uncertainties on BH mass values are rather large, see
McClintock \& Remillard 2006 and references therein) in which a type-B
QPO has been observed at 6 Hz (when its quality factor $Q>2$).
This strengthens the already known independence of this type of QPO on the
mass of the compact object (see Casella et al. 2005).

\subsection{X-ray$/$radio association} \label{subpar:disc_radio}

Klein-Wolt et al. (2002) made a detailed study of simultaneous
radio$/$X-ray observations of GRS 1915+105, focusing in particular on
radio oscillation events, and found that $\beta$ class observations
are usually associated to strong radio oscillations, while $\mu$ class
observations are usually associated to a weak steady radio
activity. These authors point out that the long hard intervals (which
are the main macroscopic difference between the two variability
classes) appear to be directly linked to the production of radio
oscillations. We found the type-B QPO in both classes $\beta$ and
$\mu$. We could not find any timing property (on time scales shorter
than 2 seconds) clearly differentiating the two classes.  This
apparently weakens the link between the presence of the type-B QPO and
radio activity. Unfortunately, the lack of radio coverage during
the analyzed RXTE observations does not allow any conclusive
result. In order to clarify this issue it will be fundamental to
extend our analysis to RXTE observations for which a radio coverage is
available.

\section*{Acknowledgments}

The authors thank Michiel van der Klis for very useful comments and
discussion. This work was supported by NWO VIDI grant to R. P. Fender and
NWO SPINOZA grant to M. B. M. van der Klis. PC acknowledges support
from an NWO Post-doctoral Fellow. This research has made use of data obtained
through the High Energy Astrophysics Science Archive Research Center Online Service,
provided by the NASA Goddard Space Flight Center.

\appendix
\section[]{Fit results and phase lags} \label{app_tables}
Here we report five tables containing the fit results for the analysed observations and two
tables with the results of the phase lag analysis. 
\newpage
\begin{table}
\centering
\caption{Fit results for Obs. \#1. We accounted for the Poissonian noise adding an additive constant to the model (see
\S \ref{par:obs_data}). In the first column the bold number indicates the region, the second number indicates the
selection within the region and it increases following the arrows in figure \ref{fig:cd_hid_arrows}. For peaks
fitted with two Lorentzians, both the components are reported. If the difference between the negative and the
positive error was smaller than 0.1, we reported just the value of the largest error (in absolute value). With ``Fixed
fit'' we refer to those cases when QPOs were detected only in one or two energy ranges: in these cases we fixed the
centroid frequency and the $FWHM$ to the values obtained from the energy range in which the peak was more significant
(see \S \ref{par:results}).} \label{tab:obs1}
\begin{tabular}{c c r r r r}
\hline
\hline
\multicolumn{6}{c}{{\bf Observation \#1}}\\
\hline
Sel. & Type   & Freq. (Hz)              & $FWHM$                 & Sign. ($\sigma$) & {\it rms ($\%$)} \\
\multicolumn{6}{c}{2-5 keV}\\
\hline
{\bf2}-1  & B      & 2.75$\pm$0.16           & 1.72$_{-0.36}^{+0.42}$ &    4.68          & 4.18$\pm$0.45    \\
{\bf2}-1  & C      & 9.35$\pm$0.33           & 1.67$_{-0.68}^{+1.07}$ &    3.10          & 2.04$\pm$0.39    \\
{\bf3}-1  & B      & 3.99$\pm$0.07           & 2.21$\pm$0.21          &    13.26         & 3.46$\pm$0.13    \\
{\bf3}-1  & C      & 10.92$\pm$0.32          & 4.33$_{-0.73}^{+0.88}$ &    7.00          & 2.05$\pm$0.16    \\
{\bf3}-2  & B      & 5.04$\pm$0.06           & 2.39$\pm$0.21          &    16.13         & 3.15$\pm$0.10    \\
{\bf3}-2  & C      & 13.09$_{-0.60}^{+0.48}$ & 4.10$_{-1.53}^{+2.12}$ &    3.45          & 1.25$\pm$0.23    \\
{\bf3}-3  & B      & 5.65$\pm$0.09           & 2.70$\pm$0.27          &    14.00         & 3.33$\pm$0.12    \\
{\bf3}-3  & C      & 12.43                   & 4.39                   &    Fixed fit     & 0.91$\pm$0.20    \\
{\bf3}-4  & B      & 6.01$\pm$0.16           & 2.68$\pm$0.49          &    8.53          & 2.06$\pm$0.13    \\
{\bf3}-4  & C      & 14.53                   & 2.66    &    Upp. lim.$^{\mathrm{d}}$     & 0.70             \\
{\bf3}-5  & B      & 5.94                    & 5.51    &    Upp. lim.$^{\mathrm{d}}$     & 1.47             \\
{\bf4}-1  & B      & 6.39$\pm$0.20           & 1.99$_{-0.62}^{+0.91}$ &    4.48          & 3.07$\pm$0.41    \\
\hline
\multicolumn{6}{c}{5-13 keV}\\
\hline
{\bf2}-1  & B      & 2.75                    & 1.72                   &    Fixed fit     & 5.23$\pm$0.52    \\
{\bf2}-1  & C      & 8.67$\pm$0.27           & 5.42$_{-0.83}^{+0.95}$ &    7.87          & 7.02$\pm$0.45    \\
{\bf3}-1  & B      & 4.27$\pm$0.04           & 1.84$\pm$0.10          &    23.23         & 5.60$\pm$0.12    \\
{\bf3}-1  & C      & 10.67$\pm$0.10          & 3.95$\pm$0.27          &    19.09         & 3.70$\pm$0.10    \\
{\bf3}-2 & B-1st   & 4.81$\pm$0.08           & 1.39$\pm$0.18          &    5.56          & -$^{\mathrm{a}}$ \\
{\bf3}-2 & B-2nd   & 5.72$\pm$0.06           & 1.03$\pm$0.11          & 5.01 & 5.23$\pm$0.50$^{\mathrm{b}}$ \\
{\bf3}-2  & C      & 12.80$\pm$0.40          & 8.92$_{-1.36}^{+1.56}$ &    6.74          & 2.40$\pm$0.18    \\
{\bf3}-3 & B-1st   & 5.28$\pm$0.12           & 1.81$\pm$0.20          &    8.08          & -$^{\mathrm{a}}$ \\ 
{\bf3}-3 & B-2nd   & 6.34$\pm$0.06           & 0.82$\pm$0.14          & 4.48 & 6.38$\pm$0.52$^{\mathrm{b}}$ \\
{\bf3}-3  & C      & 12.43$\pm$0.58          & 4.39$_{-1.52}^{+2.26}$ &    3.56          & 1.58$\pm$0.27    \\
{\bf3}-4  & B      & 6.10$\pm$0.05           & 2.32$\pm$0.15          &    21.58         & 4.20$\pm$0.10    \\
{\bf3}-4  & C      & 14.53$\pm$0.35          & 2.66$_{-0.72}^{+1.10}$ &    4.27          & 1.16$\pm$0.16    \\
{\bf3}-5  & B      & 5.94$_{-0.50}^{+0.37}$  & 5.51$_{-0.89}^{+1.12}$ &    6.39          & 3.42$\pm$0.35    \\
{\bf4}-1  & B      & 6.78$\pm$0.08           & 0.63$_{-0.21}^{+0.80}$ &    5.43   & 3.66$_{-0.34}^{+1.55}$  \\ 
\hline
\multicolumn{6}{c}{13-38 keV}\\
\hline
{\bf2}-1  & B      & 2.74                    & 1.72   &   Upp. lim.$^{\mathrm{d}}$	 & 4.38 	    \\
{\bf2}-1  & C      & 9.08$\pm$0.30           & 4.01$_{-0.84}^{+0.96}$  &   7.43 	 & 17.25$\pm$1.20   \\
{\bf3}-1  & B      & 4.27                    & 1.84                    & Fixed fit & 4.92$_{-0.50}^{+1.35}$ \\
{\bf3}-1  & C      & 11.24$\pm$0.10          & 3.53$\pm$0.39           &   10.91	 & 12.32$\pm$0.56   \\
{\bf3}-2  & B      & 5.77$\pm$0.15           & 1.37$_{-0.42}^{+0.54}$  &   3.93 	 & 4.80$\pm$0.69    \\
{\bf3}-2  & C      & 12.65$\pm$0.13          & 2.77$_{-0.48}^{+0.54}$  &   5.19 	 & 8.70$\pm$0.87    \\
{\bf3}-2 & C$^{\mathrm{c}}$ & 17.88$_{-2.77}^{+3.48}$ & 19.96$_{-7.88}^{+0.04}$ & 3.76 & 10.12$_{-1.35}^{+1.58}$ \\
{\bf3}-3  & B      & 5.65                    & 2.70                    &   Fixed fit	 & 6.82$\pm$0.50    \\
{\bf3}-3  & C      & 14.53                   & 2.66                    &   Fixed fit	 & 2.89$\pm$0.53    \\
{\bf4}-1  & B      & 6.78                    & 0.63                    &   Fixed fit	 & 7.83$\pm$0.67    \\
\hline
\hline
\end{tabular}
\begin{list}{}{}
\item[$^{\mathrm{a}}$] Peak fitted with two Lorentzians. The value of the {\it rms} is reported at the next line
\item[$^{\mathrm{b}}$] Sum of the {\it rms} of the two Lorentzians fitting this peak
\item[$^{\mathrm{c}}$] Harmonic peak
\item[$^{\mathrm{d}}$] Estimated at 3$\sigma$ confidence level
\end{list}
\end{table}
\begin{table} 
\centering
\caption{Fit results for Obs. \#2. See caption of Table \ref{tab:obs3} for an explanation.} \label{tab:obs2}
\begin{tabular}{c c r r r r}
\hline
\hline
\multicolumn{6}{c}{{\bf Observation \#2}}\\
\hline
Sel. & Type   & Freq. (Hz)              & $FWHM$                 & Sign. ($\sigma$) & {\it rms ($\%$)} \\
\multicolumn{6}{c}{2-5 keV}\\
\hline
{\bf1}-1  & C     & 2.86$\pm$0.06	    & 1.25$\pm$0.19	      &   6.46  	& 5.94$\pm$0.46     \\	  
{\bf1}-1  & C$^{\mathrm{c}}$ & 4.99$\pm$0.43 & 3.41$\pm$0.68	      &   4.68  &  4.36$_{-0.47}^{+0.69}$   \\
{\bf1}-2  & C     & 4.27$\pm$0.08	    & 1.77$\pm$0.29	      &   6.99  	& 4.54$\pm$0.42     \\
{\bf1}-3  & C     & 6.06$\pm$0.13	    & 1.74$\pm$0.33	      &   5.98  	& 3.37$\pm$0.28     \\
{\bf2}-1  & C     & 7.72$\pm$0.13	    & 1.92$\pm$0.40	      &   7.06  	& 2.77$\pm$0.20     \\
{\bf2}-2  & C     & 9.98$\pm$0.19	    & 2.98$\pm$0.55	      &   7.11  	& 2.11$\pm$0.15     \\
{\bf3}-1  & B     & 5.54$\pm$0.15	    & 1.07$_{-0.28}^{+0.38}$  &   4.68  	& 1.72$\pm$0.24     \\
{\bf3}-1  & C     & 12.57$_{-0.33}^{+0.22}$ & 1.70$_{-0.54}^{+0.95}$  &   3.01  	& 1.24$\pm$0.21     \\
{\bf3}-2  & B     & 5.10		    & 0.98		      & Fixed fit & 0.83$_{-0.23}^{+0.63}$  \\	   
{\bf3}-2  & C     & 12.72		    & 5.51     & Upp. lim.$^{\mathrm{d}}$       & 0.90              \\
\hline
\multicolumn{6}{c}{5-13 keV}\\
\hline
{\bf1}-1  & C     & 2.96$\pm$0.58	    & 1.43$\pm$0.13	      &   11.34  	& 13.36$\pm$0.60    \\
{\bf1}-1  & C$^{\mathrm{c}}$ & 5.25$\pm$0.61 & 5.44$\pm$0.61	      &   5.16  &  8.23$_{-0.81}^{+1.10}$   \\
{\bf1}-2  & C     & 4.18$\pm$0.06	    & 2.15$\pm$0.21	      &   11.76  	& 9.73$\pm$0.47     \\
{\bf1}-3  & C     & 6.06$\pm$0.07	    & 1.25$\pm$0.19	      &   9.75   	& 5.31$\pm$0.31     \\
{\bf2}-1  & C     & 7.91$\pm$0.06	    & 1.74$\pm$0.16   	      &   25.56 & 5.29$_{-0.10}^{+0.20}$    \\
{\bf2}-1 & C$^{\mathrm{c}}$ & 15.17$_{-2.18}^{+1.35}$ & 15.48$_{-3.62}^{+4.51}$ & 4.68 & 3.55$_{-0.39}^{+0.58}$    \\
{\bf2}-2  & C     & 9.81$\pm$0.08 	    & 2.41$\pm$0.26           &   14.20  	& 3.62$\pm$0.13     \\
{\bf2}-2  & C$^{\mathrm{c}}$ & 21.15$\pm$0.92 & 7.31$_{-2.32}^{+3.80}$  & 4.68  & 1.68$_{-0.18}^{+0.32}$    \\
{\bf3}-1 & B-1st  & 5.75$\pm$0.01 	    & $<$0.5$^{\mathrm{e}}$   &   8.07   	&  -$^{\mathrm{a}}$ \\
{\bf3}-1 & B-2nd  & 4.57$\pm$0.10 	    & 1.39$\pm$0.23           & 7.09 & 4.59$\pm$0.33$^{\mathrm{b}}$ \\
{\bf3}-1  & C     & 12.78$\pm$0.12	    & 2.1$\pm$50.34           &   9.09   	& 2.23$\pm$0.13     \\
{\bf3}-2  & B     & 5.10$\pm$0.10 	    & 0.98$\pm$0.21           &   7.08   	& 3.80$\pm$0.27     \\
{\bf3}-2  & C     & 11.75$\pm$0.18	    & 1.72$_{-0.38}^{+0.48}$  &   5.47   	& 1.83$\pm$0.19     \\
\hline
\multicolumn{6}{c}{13-38 keV}\\
\hline
{\bf1}-1  & C     & 2.89$\pm$0.06	    & 1.67$\pm$0.21	      &   6.56   	& 20.17$\pm$1.57    \\
{\bf1}-2  & C     & 4.01$\pm$0.17	    & 3.54$\pm$0.43	      &   6.06   	& 17.62$\pm$1.46    \\
{\bf1}-3  & C     & 6.55$\pm$0.10	    & 1.59$\pm$0.35	      &   6.02   	& 10.24$\pm$0.85    \\
{\bf1}-3 & C$^{\mathrm{c}}$ & 13.19$_{-2.67}^{+2.26}$ & 11.33$_{-3.26}^{+3.66}$ & 3.37  & 7.85$_{-1.18}^{+1.58}$    \\
{\bf2}-1  & C     & 7.97$\pm$0.13	    & 1.72$\pm$0.39	      &   4.10  & 10.74$_{-1.31}^{+1.04}$   \\
{\bf2}-2  & C     & 9.84$\pm$0.14	    & 1.73$\pm$0.35	      &   5.19  & 8.47$_{-0.82}^{+0.93}$    \\
{\bf3}-1  & B     & 5.16$\pm$0.26 	    & 1.07$_{-0.55}^{+0.68}$  &   5.45   	& 7.04$\pm$0.71     \\
{\bf3}-1  & C     & 12.94$\pm$0.18	    & 5.05$\pm$0.33	      &   10.36   	& 9.92$\pm$0.48     \\
{\bf3}-2  & B     & 4.74$\pm$0.245	    & 1.36$_{-0.21}^{+0.56}$  &   6.07  & 8.10$_{-0.67}^{+1.40}$    \\
{\bf3}-2  & C     & 12.72$_{-0.29}^{+0.43}$ & 5.51$_{-1.06}^{+0.86}$  &   7.07  & 9.92$_{-0.71}^{+0.52}$    \\
\hline
\hline
\end{tabular}
\begin{list}{}{}
\item[$^{\mathrm{a}}$] Peak fitted with two Lorentzians. The value of the {\it rms} is reported at the next line
\item[$^{\mathrm{b}}$] Sum of the {\it rms} of the two Lorentzians fitting this peak
\item[$^{\mathrm{c}}$] Harmonic peak
\item[$^{\mathrm{d}}$] Estimated at 3$\sigma$ confidence level
\item[$^{\mathrm{e}}$] Our resolution is 0.5 s (we extracted power density spectra on 2-second intervals, see
\S \ref{par:obs_data})
\end{list}
\end{table}
\begin{table*} 
\centering
\caption{Fit results for Obs. \#3. In the first column the bold number indicates the region, the second number
indicates the selection within the region and it increases following the arrows in figure \ref{fig:cd_hid_arrows}.
For peaks
fitted with two Lorentzians, both the components are reported. If the difference between the negative and the
positive error was smaller than 0.1, we reported just the value of the largest error (in absolute value).
With ``Fixed
fit'' we refer to those cases when QPOs were detected only in one or two energy ranges: in these cases we fixed the
centroid frequency and the $FWHM$ to the values obtained from the energy range in which the peak was more significant
(see \S \ref{par:results}).} \label{tab:obs3}
\begin{tabular}{c c r r r r}
\hline
\hline
\multicolumn{6}{c}{{\bf Observation \#3}}\\
\hline
Sel. & Type   & Freq. (Hz)              & $FWHM$                 & Sign. ($\sigma$) & {\it rms ($\%$)} \\
\multicolumn{6}{c}{2-5 keV}\\
\hline
{\bf1}-1  & C     & 2.72                    & 2.29                    &   Fixed fit	& 10.12$\pm$0.35    \\ 
{\bf1}-2  & C     & 3.58$\pm$0.09           & 2.41$\pm$0.28           &   6.57          & 6.42$\pm$0.53     \\
{\bf1}-3  & C     & 5.80$\pm$0.10           & 1.93$\pm$0.28           &   8.66          & 4.36$\pm$0.26     \\
{\bf1}-3 & C$^{\mathrm{c}}$ & 13.41$_{-0.61}^{+0.71}$ & 5.60$_{-1.98}^{+3.28}$ & 3.74 & 2.48$_{-0.33}^{+0.45}$ \\
{\bf2}-1  & B     & 2.44$\pm$0.28           & 2.24$\pm$0.49           &   3.06	& 2.95$_{-0.48}^{+0.68}$    \\
{\bf2}-1  & C     & 7.57$\pm$0.07           & 2.28$\pm$0.24           &   21.08         & 3.14$\pm$0.13     \\
{\bf2}-1 & C$^{\mathrm{c}}$ & 15.91$_{-4.00}^{+3.26}$ & 19.98$_{-1.42}^{+0.02}$ & 5.01 & 2.01 $_{-0.20}^{+0.30}$ \\
{\bf3}-1  & B     & 5.03$\pm$0.33           & 2.62$_{-0.61}^{+0.92}$  &   4.00 &  1.77$_{-0.22}^{+0.33}$    \\
{\bf3}-1  & C     & 11.00$\pm$0.21          & 3.66$_{-0.57}^{+0.69}$  &   6.94          & 1.84$\pm$0.13     \\
{\bf3}-2  & B     & 5.91$\pm$0.09           & 0.71$\pm$0.18           &   5.46          & 1.47$\pm$0.13     \\
{\bf3}-2  & C     & 12.59$\pm$0.27          & 3.69$\pm$0.84           &   6.79          & 1.70$\pm$0.15     \\
{\bf3}-3  & B     & 5.53                    & 0.75                    &   Fixed fit	& 0.41$\pm$0.28     \\
{\bf3}-4$^{\mathrm{a}}$  & B     & 5.37     & 1.45                    &   Fixed fit	& 0.51$\pm$0.10     \\
\hline
\multicolumn{6}{c}{5-13 keV}\\
\hline
{\bf1}-1  & C     & 2.72$\pm$0.14           & 4.01$\pm$0.23           &   15.77         & 18.04$\pm$0.58    \\
{\bf1}-2  & C     & 3.35$\pm$0.08           & 2.80$\pm$0.22           &   8.23 &  14.13$_{-0.86}^{+1.03}$   \\
{\bf1}-3  & C     & 5.71$\pm$0.09           & 2.13$\pm$0.21           &   9.09          & 8.61$\pm$0.47     \\
{\bf1}-3  & C$^{\mathrm{c}}$ & 13.90$\pm$0.54 & 10.36$_{-1.10}^{+1.23}$  &   9.81   	& 5.58$\pm$0.31     \\
{\bf2}-1  & B     & 2.64$\pm$0.21           & 2.24$\pm$0.38           &   3.50 &  4.40$_{-0.63}^{+0.76}$    \\
{\bf2}-1  & C     & 7.65$\pm$0.04           & 1.94$\pm$0.10           &   24.82  	& 5.74$\pm$0.12     \\
{\bf2}-1  & C$^{\mathrm{c}}$ & 14.85$_{-0.92}^{+0.77}$ & 15.89$_{-1.66}^{+1.82}$ & 9.05 & 3.88$\pm$0.25     \\
{\bf3}-1 & B-1st  & 5.77$\pm$0.03           & $<$0.5$^{\mathrm{f}}$   &   6.93   	&  -$^{\mathrm{b}}$ \\
{\bf3}-1 & B-2nd & 4.17$\pm$0.12            & 2.31$\pm$0.41           & 5.42 & 4.04$\pm$0.43$^{\mathrm{d}}$ \\
{\bf3}-1  & C     & 11.17$\pm$0.12          & 3.89$\pm$0.29           &   15.98         & 3.13$\pm$0.10     \\ 
{\bf3}-1 & -$^{\mathrm{e}}$ & 25.00$_{-1.43}^{+1.30}$ & 9.37$_{-2.23}^{+3.35}$ &   3.80 & 1.27$\pm$0.20     \\ 
{\bf3}-2 & B-1st  & 5.17$\pm$0.06           & 1.00$\pm$0.14           &   13.78         &  -$^{\mathrm{b}}$ \\
{\bf3}-2 & B-2nd  & 6.25$\pm$0.02           & $<$0.5$^{\mathrm{f}}$   & 8.23 & 3.97$\pm$0.18$^{\mathrm{d}}$ \\
{\bf3}-2  & C     & 13.19$\pm$0.16          & 3.66$\pm$0.45           &   13.39         & 2.41$\pm$0.11     \\
{\bf3}-3  & B     & 5.53$\pm$0.05           & 0.75$\pm$0.23           &   8.14   	& 1.99$\pm$0.13     \\
{\bf3}-4$^{\mathrm{a}}$ & B & 5.36$\pm$0.16 & 1.45$_{-0.29}^{+0.45}$  &   3.90   	& 0.78$\pm$0.10     \\
\hline
\multicolumn{6}{c}{13-38 keV}\\
\hline
{\bf1}-1  & C     & 2.80$\pm$0.22           & 3.42$\pm$0.36           &   8.13 	& 22.86$_{-1.42}^{+1.54}$   \\
{\bf1}-2  & C     & 3.20$\pm$0.10           & 3.18$\pm$0.20           &   11.42         & 21.77$\pm$0.96    \\
{\bf1}-3  & C     & 6.15$\pm$0.07           & 1.29$\pm$0.30           &   5.95   	& 10.61$\pm$0.89    \\
{\bf2}-1  & B     & 2.64                    & 2.24                    &   Fixed fit 	& 5.29$\pm$0.69     \\
{\bf2}-1  & C     & 7.71$\pm$0.06           & 1.58$\pm$0.14           &   9.84          & 11.65$\pm$0.59    \\
{\bf2}-1 & C$^{\mathrm{c}}$ & 17.72$_{-2.27}^{+1.92}$ & 15.29$_{-3.53}^{+3.97}$ & 3.55 & 7.37$_{-1.05}^{+0.92}$ \\ 
{\bf3}-1  & B     & 5.29$\pm$0.21           & 1.61$_{-0.44}^{+0.65}$  &   4.32  &  4.65$_{-0.54}^{+0.75}$   \\
{\bf3}-1  & C     & 11.35$\pm$0.11          & 4.21$\pm$0.44           &   7.80  &  11.05$_{-0.71}^{+0.45}$  \\
{\bf3}-1 & C$^{\mathrm{c}}$ & 20.37$_{-4.57}^{+3.50}$ & 19.95$_{-5.79}^{+0.05}$ & 3.93  &  6.48$_{-0.86}^{+1.41}$ \\
{\bf3}-2  & B     & 5.38$\pm$0.08           & 1.47$\pm$0.19           &   10.34         & 8.39$\pm$0.43     \\
{\bf3}-2  & C     & 12.57$\pm$0.19          & 2.98$\pm$0.54           &   4.02  &  7.73$_{-0.96}^{+0.86}$   \\
{\bf3}-3  & B     & 4.72$\pm$0.47           & 3.97$_{-1.07}^{+1.78}$  &   4.14  &  5.47$_{-0.67}^{+0.88}$   \\
{\bf3}-4$^{\mathrm{a}}$ & B & 4.91$_{-0.53}^{+0.33}$ & 2.80$_{-1.04}^{+2.69}$ & 3.44 &  2.37$_{-0.35}^{+0.70}$   \\
\hline
\hline
\end{tabular}
\begin{list}{}{}
\item[$^{\mathrm{a}}$] Poissonian noise fitted adding a constant to the model, since the Poissonian noise
subtraction left flat residuals at high frequencies (see \S \ref{par:obs_data})
\item[$^{\mathrm{b}}$] Peak fitted with two Lorentzians. The value of the {\it rms} is reported at the next line
\item[$^{\mathrm{c}}$] Harmonic peak
\item[$^{\mathrm{d}}$] Sum of the {\it rms} of the two Lorentzians fitting this peak
\item[$^{\mathrm{e}}$] High frequency QPO, not considered in our analysis (see Belloni et al. 2006)
\item[$^{\mathrm{f}}$] Our resolution is 0.5 s (we extracted power density spectra on 2-second intervals, see
\S \ref{par:obs_data})
\end{list}
\end{table*}
\begin{table} 
\centering
\caption{Fit results for Obs. \#4. See caption of Table \ref{tab:obs3} for an explanation.} \label{tab:obs4}
\begin{tabular}{c c r r r r}
\hline
\hline
\multicolumn{6}{c}{{\bf Observation \#4}}\\
\hline
Sel.      & Type   & Freq. (Hz)              & $FWHM$                 & Sign. ($\sigma$) & {\it rms ($\%$)} \\
\multicolumn{6}{c}{2-5 keV}\\
\hline
{\bf2}-1  & C      & 10.32                   & 3.46                   &    Fixed fit     & 2.19$\pm$0.27    \\
{\bf3}-1  & B      & 4.11$\pm$0.07           & 2.13$\pm$0.23          &    13.50         & 3.21$\pm$0.13    \\
{\bf3}-1  & C      & 11.91$\pm$0.21          & 3.15$_{-0.55}^{+0.71}$ &    8.88          & 1.81$\pm$0.16    \\
{\bf3}-2  & B      & 5.04$\pm$0.06           & 1.96$\pm$0.18          &    13.76         & 2.78$\pm$0.10    \\
{\bf3}-2  & C      & 11.91$_{-0.52}^{+0.42}$ & 4.37$_{-1.26}^{+2.01}$ &    3.85          & 1.41$\pm$0.18    \\
{\bf3}-3  & B      & 5.65$\pm$0.12           & 2.70$_{-0.28}^{+0.39}$ &    10.44         & 2.96$\pm$0.17    \\
{\bf3}-3  & C      & 13.63                   & 6.72                   &    Fixed fit     & 1.02$\pm$0.22    \\
{\bf3}-4  & B      & 5.56$\pm$0.09           & 0.99$\pm$0.24          &    4.54          & 1.64$\pm$0.18    \\ 
{\bf4}-1  & B      & 6.84$\pm$0.15           & 2.06$\pm$0.42          &    5.24          & 3.17$\pm$0.02    \\
{\bf4}-1 & B$^{\mathrm{c}}$ & 12.00$_{-1.19}^{+0.97}$ & 7.98$_{-2.12}^{+3.95}$ & 3.70 & 2.73$_{-0.34}^{+0.46}$    \\ 
\hline
\multicolumn{6}{c}{5-13 keV}\\
\hline
{\bf2}-1  & C      & 9.92$\pm$0.21           & 3.09$\pm$0.50          &    6.70          & 4.80$\pm$0.36    \\
{\bf3}-1  & B      & 4.27$\pm$0.04           & 2.01$\pm$0.10          &    24.77         & 6.20$\pm$0.13    \\
{\bf3}-1  & C      & 11.45$\pm$0.11          & 4.34$\pm$0.39          &    18.25         & 3.71$\pm$0.12    \\
{\bf3}-2  & B-1st  & 4.82$\pm$0.06           & 1.38$\pm$0.12          &    10.15   	 & -$^{\mathrm{a}}$ \\ 
{\bf3}-2  & B-2nd  & 5.79$\pm$0.05           & 0.66$\pm$0.12          & 4.75 & 5.93$\pm$0.39$^{\mathrm{b}}$ \\ 
{\bf3}-2  & C      & 12.52$\pm$0.57          & 9.94$_{-1.56}^{+1.85}$ &    7.49   	 & 2.54$\pm$0.19    \\
{\bf3}-3  & B      & 5.79$\pm$0.05           & 1.61$\pm$0.10          &    22.08         & 5.87$\pm$0.14    \\
{\bf3}-3  & C      & 13.63                   & 6.72                   &    Fixed fit     & 0.97$\pm$0.27    \\
{\bf3}-4  & B      & 5.61$\pm$0.05           & 1.74$\pm$0.16          &    16.41         & 4.12$\pm$0.13    \\
{\bf3}-4  & B$^{\mathrm{c}}$  & 11.13$_{-0.76}^{+0.62}$ & 5.65$_{-1.23}^{+2.20}$ & 3.89  & 1.63$\pm$0.27    \\
{\bf4}-1  & B      & 6.79$\pm$0.07           & 1.66$\pm$0.22          &    10.79         & 6.48$\pm$0.30    \\
{\bf4}-1  & B$^{\mathrm{c}}$ & 13.46$_{-1.49}^{+1.10}$ & 14.54$_{-2.13}^{+2.54}$ & 6.68  & 5.54$\pm$0.34    \\
\hline
\multicolumn{6}{c}{13-38 keV}\\
\hline
{\bf2}-1  & C      & 10.32$\pm$0.19          & 3.46$_{-0.53}^{+0.63}$ &    9.84          & 13.68$\pm$0.71   \\
{\bf3}-1  & B      & 4.27                    & 2.01                   &    Fixed fit     & 4.02$\pm$0.99    \\ 
{\bf3}-1  & C      & 11.67$\pm$0.13          & 2.82$\pm$0.44          &    6.90          & 9.49$\pm$0.69    \\
{\bf3}-2  & B      & 4.62$\pm$0.31           & 2.33$_{-0.73}^{+0.96}$ &    3.46 &  5.28$_{-0.77}^{+1.00}$   \\
{\bf3}-2  & C      & 12.67$\pm$0.19          & 5.91$\pm$0.76          &    13.31         & 11.16$\pm$0.45   \\
{\bf3}-3  & B      & 5.85$\pm$0.30           & 2.39$\pm$0.63          &    4.01  &  7.20$_{-0.90}^{+0.65}$  \\
{\bf3}-3  & C      & 13.63$\pm$0.54          & 6.72$_{-1.67}^{+2.66}$ &    6.39  &  8.97$_{-0.94}^{+1.68}$  \\
{\bf3}-4  & B      & 5.41$_{-0.79}^{+0.62}$  & 6.01$_{-2.24}^{+3.50}$ &    4.66  &  5.74$_{-0.65}^{+1.54}$  \\
{\bf4}-1  & B      & 6.83$\pm$0.14           & 1.52$_{-0.35}^{+0.45}$ &    5.28  &  10.76$_{-1.02}^{+1.78}$ \\
\hline
\hline
\end{tabular}
\begin{list}{}{}
\item[$^{\mathrm{a}}$] Peak fitted with two Lorentzians. The value of the {\it rms} is reported at the next line
\item[$^{\mathrm{b}}$] Sum of the {\it rms} of the two Lorentzians fitting this peak
\item[$^{\mathrm{c}}$] Harmonic peak
\end{list}
\end{table}
\begin{table} 
\centering
\caption{Fit results for Obs. \#5. See caption of Table \ref{tab:obs3} for an explanation.} \label{tab:obs5}
\begin{tabular}{c c r r r r}
\hline
\hline
\multicolumn{6}{c}{{\bf Observation \#5}}\\
\hline
Sel.      & Type   & Freq. (Hz)              & $FWHM$                 & Sign. ($\sigma$) & {\it rms ($\%$)} \\
\multicolumn{6}{c}{2-5 keV}\\
\hline
{\bf1}-1  & C      & 2.53$\pm$0.22           & 3.92$\pm$0.41          &    7.99   & 9.20$_{-0.56}^{+0.37}$  \\
{\bf1}-2  & C      & 3.17$\pm$0.20           & 4.61$\pm$0.34          &    9.50          & 8.44$\pm$0.45    \\
{\bf1}-3  & C      & 6.42$\pm$0.05           & 1.26$\pm$0.13          &    14.33         & 3.28$\pm$0.13    \\
{\bf1}-3  & C$^{\mathrm{c}}$ & 14.03$\pm$0.65 & 8.58$_{-1.74}^{+2.38}$ & 6.46            & 2.50$\pm$0.27    \\
{\bf2}-1  & B      & 2.75$\pm$0.10           & 2.50$\pm$0.39          &    13.13         & 2.87$\pm$0.13    \\
{\bf2}-1  & C      & 8.40$\pm$0.16           & 3.14$\pm$0.40          &    4.42          & 2.53$\pm$0.38    \\
{\bf3}-1  & B      & 3.99$_{-0.52}^{+0.38}$  & 5.14$_{-0.73}^{+0.83}$ &    4.80 	 & 3.41$\pm$0.36    \\
{\bf3}-1  & C      & 11.88$\pm$0.15          & 1.71$_{-0.50}^{+0.66}$ &    3.39  	 & 1.25$\pm$0.18    \\
{\bf3}-2  & B      & 5.59$\pm$0.20           & 2.59$\pm$0.39          &    7.71  	 & 3.09$\pm$0.23    \\
{\bf3}-2  & C      & 12.85$\pm$0.33          & 2.97$_{-0.61}^{+0.73}$ &    5.66  	 & 1.69$\pm$0.16    \\
{\bf3}-3  & B      & 5.02$\pm$0.31           & 2.28$_{-0.72}^{+1.11}$ &    3.28  	 & 2.26$\pm$0.41    \\
{\bf3}-3  & C      & 15.31                   & 6.22                   &    Fixed fit     & 1.46$\pm$0.42    \\
{\bf4}-1  & B      & 5.96$\pm$0.12           & 0.76$_{-0.24}^{+0.36}$ &    3.50  	 & 1.84$\pm$0.32    \\
\hline
\multicolumn{6}{c}{5-13 keV}\\
\hline
{\bf1}-1  & C-1st  & 2.70$_{-0.14}^{+0.03}$  & $<$0.5$^{\mathrm{e}}$  &    3.29   	 & -$^{\mathrm{a}}$ \\
{\bf1}-1  & C-2nd  & 2.69$\pm$0.27          & 4.27$\pm$0.30 & 6.26 & 17.88$_{-1.53}^{+1.44}$$^{\mathrm{b}}$ \\
{\bf1}-2  & C      & 3.55$\pm$0.21           & 3.05$\pm$0.44          &    4.71  &  13.58$_{-1.45}^{+2.16}$ \\
{\bf1}-3  & C      & 6.46$\pm$0.03           & 1.58$\pm$0.08          &    26.59   	 & 7.60$\pm$0.14    \\
{\bf1}-3  & C$^{\mathrm{c}}$ & 12.86$_{-0.42}^{+0.22}$ & 12.36$\pm$0.93 &    13.58   	 & 5.87$\pm$0.24    \\
{\bf2}-1  & B      & 2.84$\pm$0.12           & 2.53$\pm$0.26          &    6.89   	 & 4.70$\pm$0.39    \\
{\bf2}-1  & C      & 8.70$\pm$0.05           & 3.92$\pm$0.17          &    26.56   	 & 5.63$\pm$0.11    \\
{\bf2}-1  & C$^{\mathrm{c}}$ & 18.43$\pm$0.59 & 11.55$_{-1.45}^{+1.66}$ &   8.58   	 & 2.94$\pm$0.20    \\
{\bf3}-1  & B-1st  & 4.40$\pm$0.12           & 2.85$\pm$0.15          &    10.30   	 & -$^{\mathrm{a}}$ \\ 
{\bf3}-1  & B-2nd  & 6.25$\pm$0.005          & $<$0.5$^{\mathrm{e}}$  & 7.94 & 5.54$\pm$0.29$^{\mathrm{b}}$ \\
{\bf3}-1  & C      & 12.08$\pm$0.07          & 2.27$\pm$0.22          &    15.28   	 & 2.49$\pm$0.08    \\
{\bf3}-2  & B-1st  & 5.07$\pm$0.08           & 1.62$\pm$0.30          &    6.57   	 & -$^{\mathrm{a}}$ \\
{\bf3}-2  & B-2nd  & 6.42$\pm$0.10           & $<$0.5$^{\mathrm{e}}$  & 3.50 & 5.52$\pm$0.50$^{\mathrm{b}}$ \\
{\bf3}-2  & C      & 13.13$\pm$0.25          & 4.97$_{-0.69}^{+0.79}$ &    9.96   	 & 2.93$\pm$0.18    \\
{\bf3}-3  & B      & 5.40$\pm$0.07           & 1.56$\pm$0.16          &    12.70  	 & 4.87$\pm$0.19    \\
{\bf3}-3  & C      & 15.31$\pm$0.46          & 6.22$_{-1.11}^{+1.33}$ &    6.13   	 & 2.28$\pm$0.19    \\
{\bf4}-1  & B      & 6.10$\pm$0.10           & 1.85$\pm$0.55          &    3.46    & 4.30$_{-0.62}^{+0.97}$ \\ 
\hline
\multicolumn{6}{c}{13-38 keV}\\
\hline
{\bf1}-1  & C      & 2.80$\pm$0.11           & 2.06$_{-0.47}^{+0.33}$ &    3.21   & 19.10$_{-2.99}^{+1.24}$ \\
{\bf1}-2  & C      & 3.21$\pm$0.22           & 3.84$\pm$0.35          &    5.81   & 22.67$_{-1.96}^{+1.62}$ \\
{\bf1}-3  & C      & 6.69$\pm$0.04           & 1.76$\pm$0.14          &    13.39  	 & 14.15$\pm$0.53   \\
{\bf1}-3  & C$^{\mathrm{c}}$  & 10.67$_{-1.07}^{+0.86}$ & 9.91$_{-1.53}^{+1.68}$ & 5.47   & 10.00$_{-0.92}^{+1.25}$ \\
{\bf2}-1  & B      & 2.84                    & 2.53       & Upp. lim.$^{\mathrm{d}}$     & 1.83             \\
{\bf2}-1  & C-1st  & 10.17$\pm$0.17          & 2.67$\pm$0.33          &    5.81  	 & -$^{\mathrm{a}}$ \\
{\bf2}-1  & C-2nd  & 8.34$\pm$0.09          & 1.40$\pm$0.29 & 4.31 & 11.92$_{-1.15}^{+1.28}$$^{\mathrm{b}}$ \\
{\bf3}-1  & B      & 5.14$\pm$0.29           & 2.54$_{-0.58}^{+0.68}$ &    4.43    & 6.23$_{-0.71}^{+0.88}$ \\
{\bf3}-1  & C      & 11.81$\pm$0.09          & 2.75$\pm$0.34          &    8.80  	 & 9.18$\pm$0.55    \\
{\bf3}-2  & B      & 6.04$\pm$0.11           & 0.97$\pm$0.26          &    4.57  	 & 6.58$\pm$0.72    \\
{\bf3}-2  & C      & 13.56$\pm$0.23          & 4.27$_{-0.61}^{+0.71}$ &    10.04 	 & 9.42$\pm$0.49    \\
{\bf3}-3  & B      & 5.35$\pm$0.18           & 2.22$_{-0.50}^{+0.63}$ &    5.53    & 7.06$_{-0.64}^{+0.70}$ \\
{\bf3}-3  & C      & 17.65$_{-1.64}^{+1.35}$ & 11.76$_{-2.94}^{+4.12}$ &   4.86    & 7.30$_{-0.77}^{+0.93}$ \\
{\bf4}-1  & B      & 5.96                    & 0.76                   &  Fixed fit & 4.02$_{-1.01}^{+0.87}$ \\
\hline
\hline
\end{tabular}
\begin{list}{}{}
\item[$^{\mathrm{a}}$] Peak fitted with two Lorentzians. The value of the {\it rms} is reported at the next line
\item[$^{\mathrm{b}}$] Sum of the {\it rms} of the two Lorentzians fitting this peak
\item[$^{\mathrm{c}}$] Harmonic peak
\item[$^{\mathrm{d}}$] Estimated at 3$\sigma$ confidence level
\item[$^{\mathrm{e}}$] Our resolution is 0.5 s (we extracted power density spectra on 2-second intervals, see
\S \ref{par:obs_data})
\end{list}
\end{table}
\begin{table} 
\centering
\caption{Phase lags for each component detected in the fits for observations \# 1, 2 and 3, integrated in a range centered on the peak centroid
frequency with a width equal to the $FWHM$ of the peak itself. Centroid frequency values were taken from fits in the
5-13 keV energy band. For peaks fitted with two Lorentzians, we calculated the lag integrating in a range centered on
the component with the highest fractional {\it rms} (see \S \ref{par:obs_data}).} \label{tab:lagall1}
\begin{tabular}{c c r r}
\hline
\hline
\multicolumn{4}{c}{{\bf Observation \#1}}\\
\hline
Sel.      & Type               & Freq. (Hz)              & Phase lag (rad)   \\
\hline
{\bf2}-1  & B                  & 2.75                    & -0.042$\pm$0.020  \\
{\bf2}-1  & C                  & 8.67$\pm$0.27           & 0.051$\pm$0.019   \\ 
{\bf3}-1  & B                  & 4.27$\pm$0.04           & -0.110$\pm$0.007  \\
{\bf3}-1  & C                  & 10.67$\pm$0.10          & 0.040$\pm$0.007   \\
{\bf3}-2  & B$^{\mathrm{a}}$   & 4.81$\pm$0.08           & -0.142$\pm$0.008  \\
{\bf3}-2  & C                  & 12.80$\pm$0.40          & 0.001$\pm$0.005   \\
{\bf3}-3  & B$^{\mathrm{a}}$   & 5.28$\pm$0.12           & -0.089$\pm$0.001  \\
{\bf3}-3  & C                  & 12.43$\pm$0.58          & -0.019$\pm$0.001  \\
{\bf3}-4  & B                  & 6.10$\pm$0.05           & -0.025$\pm$0.010  \\
{\bf3}-4  & C                  & 14.53$\pm$0.35          & -0.008$\pm$0.014  \\
{\bf3}-5  & B                  & 5.94$_{-0.50}^{+0.37}$  & -0.019$\pm$0.013  \\
{\bf4}-1  & B                  & 6.78$\pm$0.08           & -0.095$\pm$0.031  \\
\hline
\multicolumn{4}{c}{{\bf Observation \#2}}\\
\hline
{\bf1}-1  & C                  & 2.96$\pm$0.58	         & -0.018$\pm$0.022  \\
{\bf1}-1  & C$^{\mathrm{d}}$   & 5.25$\pm$0.61	         & -0.023$\pm$0.020  \\
{\bf1}-2  & C                  & 4.18$\pm$0.06	         & -0.120$\pm$0.020  \\
{\bf1}-3  & C                  & 6.06$\pm$0.07	         & -0.271$\pm$0.040  \\
{\bf2}-1  & C                  & 7.91$\pm$0.06	         & -0.293$\pm$0.055  \\
{\bf2}-1  & C$^{\mathrm{d}}$   & 15.17$_{-2.18}^{+1.35}$ & -0.285$\pm$0.046  \\
{\bf2}-2  & C                  & 9.81$\pm$0.08 	         & -0.115$\pm$0.044  \\ 
{\bf2}-2  & C   & 21.15$\pm$0.92	         & -0.290$\pm$0.155  \\ 
{\bf3}-1  & B$^{\mathrm{a}}$   & 4.57$\pm$0.10 	         & -0.052$\pm$0.026  \\
{\bf3}-1  & C                  & 12.78$\pm$0.12	         & -0.123$\pm$0.066  \\
{\bf3}-2  & B                  & 5.10$\pm$0.10 	         & 0.140$\pm$0.060   \\
{\bf3}-2  & C                  & 11.75$\pm$0.18	         & -0.263$\pm$0.103  \\
\hline
\multicolumn{4}{c}{{\bf Observation \#3}}\\
\hline
{\bf1}-1  & C                  & 2.72$\pm$0.14           & -0.122$\pm$0.025  \\
{\bf1}-2  & C                  & 3.35$\pm$0.08           & -0.093$\pm$0.013  \\
{\bf1}-3  & C                  & 5.71$\pm$0.09           & -0.116$\pm$0.028  \\
{\bf1}-3  & C$^{\mathrm{d}}$   & 13.90$\pm$0.54          & -0.031$\pm$0.053  \\
{\bf2}-1  & B                  & 2.64$\pm$0.21           & -0.128$\pm$0.020  \\
{\bf2}-1  & C                  & 7.65$\pm$0.04           & -0.189$\pm$0.024  \\
{\bf2}-1  & C$^{\mathrm{d}}$   & 14.85$_{-0.92}^{+0.77}$ & -0.202$\pm$0.021  \\
{\bf3}-1  & B$^{\mathrm{a}}$   & 4.17$\pm$0.12           & -0.101$\pm$0.019  \\
{\bf3}-1  & C                  & 11.17$\pm$0.11          & -0.115$\pm$0.029  \\
{\bf3}-1  & -$^{\mathrm{b}}$   & 25.00$_{-1.43}^{+1.30}$ & -0.168$\pm$0.194  \\
{\bf3}-2  & B                  & 5.17$\pm$0.06           & 0.080$\pm$0.040   \\
{\bf3}-2  & C                  & 13.19$\pm$0.16          & -0.080$\pm$0.043  \\ 
{\bf3}-3  & B                  & 5.53$\pm$0.05           & -0.282$\pm$0.120  \\ 
{\bf3}-4$^{\mathrm{c}}$ & B    & 5.36$\pm$0.16           & 0.008$\pm$0.060   \\ 
\hline
\hline
\end{tabular}
\begin{list}{}{}
\item[$^{\mathrm{a}}$] Peak fitted with two Lorentzians
\item[$^{\mathrm{b}}$] High frequency QPO
\item[$^{\mathrm{c}}$] Poissonian noise fitted adding a constant to the model
\item[$^{\mathrm{d}}$] Harmonic peak
\end{list}
\end{table}
\begin{table} 
\centering
\caption{Phase lags for each component detected in the fits for observations \# 4 and 5. See caption of Table \ref{tab:lagall1}
for an explanation.} \label{tab:lagall2}
\begin{tabular}{c c r r}
\hline
\hline
\multicolumn{4}{c}{{\bf Observation \#4}}\\
\hline
Sel.      & Type               & Freq. (Hz)              & Phase lag (rad)   \\
\hline
{\bf2}-1  & C                  & 9.92$\pm$0.21           & 0.021$\pm$0.102   \\
{\bf3}-1  & B                  & 4.27$\pm$0.04           & -0.262$\pm$0.024  \\
{\bf3}-1  & C                  & 11.45$\pm$0.11          & 0.0$\pm$0.052     \\
{\bf3}-2  & B$^{\mathrm{a}}$   & 4.82$\pm$0.06           & -0.313$\pm$0.024  \\ 
{\bf3}-2  & C                  & 12.52$\pm$0.57          & -0.188$\pm$0.046  \\
{\bf3}-3  & B                  & 5.79$\pm$0.05           & -0.127$\pm$0.038  \\
{\bf3}-3  & C                  & 13.63$^{\mathrm{b}}$    & -0.087$\pm$0.102  \\
{\bf3}-4  & B                  & 5.61$\pm$0.05           & -0.122$\pm$0.041  \\
{\bf3}-4  & B$^{\mathrm{c}}$   & 11.13$_{-0.76}^{+0.62}$ & -0.160$\pm$0.096  \\
{\bf4}-1  & B                  & 6.79$\pm$0.07           & -0.074$\pm$0.048  \\
{\bf4}-1  & B$^{\mathrm{c}}$   & 13.46$_{-1.49}^{+1.10}$ & -0.171$\pm$0.040  \\
\hline
\multicolumn{4}{c}{{\bf Observation \#5}}\\
\hline
{\bf1}-1  & C$^{\mathrm{a}}$   & 2.69$\pm$0.27           & -0.130$\pm$0.023  \\
{\bf1}-2  & C                  & 3.55$\pm$0.21           & -0.116$\pm$0.018  \\
{\bf1}-3  & C                  & 6.46$\pm$0.03           & -0.125$\pm$0.024  \\
{\bf1}-3  & C$^{\mathrm{c}}$   & 12.86$_{-0.42}^{+0.22}$ & -0.060$\pm$0.025  \\
{\bf2}-1  & B                  & 2.84$\pm$0.12           & -0.102$\pm$0.014  \\
{\bf2}-1  & C                  & 8.70$\pm$0.05           & -0.061$\pm$0.019  \\
{\bf2}-1  & C$^{\mathrm{c}}$   & 18.43$\pm$0.59          & -0.057$\pm$0.064  \\
{\bf3}-1  & B$^{\mathrm{a}}$   & 4.40$\pm$0.12           & -0.140$\pm$0.015  \\ 
{\bf3}-1  & C                  & 12.08$\pm$0.07          & 0.033$\pm$0.046   \\
{\bf3}-2  & B$^{\mathrm{a}}$   & 5.07$\pm$0.08           & -0.267$\pm$0.036  \\
{\bf3}-2  & C                  & 13.13$\pm$0.25          & -0.104$\pm$0.052  \\
{\bf3}-3  & B                  & 5.40$\pm$0.07           & -0.290$\pm$0.050  \\
{\bf3}-3  & C                  & 15.31$\pm$0.46          & -0.029$\pm$0.073  \\
{\bf4}-1  & B                  & 6.10$\pm$0.10           & -0.0$\pm$0.044    \\ 
\hline
\hline
\end{tabular}
\begin{list}{}{}
\item[$^{\mathrm{a}}$] Peak fitted with two Lorentzians
\item[$^{\mathrm{b}}$] Fixed fit (see caption of table \ref{tab:obs1} for an explanation)
\item[$^{\mathrm{c}}$] Harmonic peak
\end{list}
\end{table}
\label{lastpage}

\end{document}